# On the thermoelectric figure of merit limits of superlattice materials

P.V.Gorsky

Institute of Thermoelectricity of the NAS and MES of Ukraine

gena.grim@gmail.com

The hopes for improvement of the thermoelectric figure of merit of superlattice materials are mainly pinned on the reduction of free charge carrier gas dimensionality in thermoelectric material. It is assumed that this reduction results in the increase of charge carrier density of states on chemical potential level, and, hence, in the increase of the Seebeck coefficient value and the thermoelectric figure of merit. However, ideally "two-dimensional" or "one-dimensional" systems do not exist, making us to study this issue in more detail.

With this object in mind, let us consider a thermoelectric material with a strongly anisotropic carrier spectrum of the type [1]:

$$\varepsilon(\vec{k}) = \frac{\hbar^2}{2m^*}(k_x^2 + k_y^2) + \Delta(1 - \cos ak_z), \qquad (1)$$

where $k_x, k_y$ – quasi-pulse components in superlattice layer plane, $k_z$ – quasi-pulse component perpendicular to layers, $m^*$ – effective mass of carriers in layer plane, $\Delta$ – half-width of miniband defining motion of free carriers along the superlattice axis, $a$ – the distance between translation equivalent layers.

In parallel with model (1) we will consider the so-called "ideal nonparabolicity model". In the framework of this model, a miniband defining free charge carrier motion along the superlattice axis has a finite width, the same as in model (1), but the parabolic dispersion law is preserved up to the boundaries of one-dimensional Brillouin zone:

$$\varepsilon(\vec{k}) = \frac{\hbar^2}{2m^*}(k_x^2 + k_y^2) + 2\Delta \frac{a^2 k_z^2}{\pi^2}. \qquad (2)$$

We now introduce the value that will be called the superlattice quasi-two-dimensionality degree, i.e. the ratio of $K$ Fermi energy $\zeta_{02D}$ of ideal two-dimensional Fermi-gas with a parabolic dispersion law to miniband width $\Delta$. Then the case $K \ll 1$ will correspond to three-dimensional free charge carrier gas with a quadratic dispersion law. The opposite case $K \gg 1$ corresponds to a quasi-two-dimensional superlattice.

Taking into account that $\zeta_{02D} = h^2 n_0 a / 4\pi m^*$, where $n_0$ – volumetric concentration of free charge carriers in material, the equations defining chemical potential $\zeta$ of free charge carrier gas in models (1) and (2), respectively, acquire the form:

$$\frac{t_{2D}}{\pi}\int_0^\pi \ln\left[1 + \exp\left(\frac{\gamma^* - K^{-1}(1 - \cos x)}{t_{2D}}\right)\right] - 1 = 0, \qquad (3)$$

$$\frac{t_{2D}}{\pi}\int_0^\pi \ln\left[1 + \exp\left(\frac{\gamma^* - 2\pi^{-2}K^{-1}x^2}{t_{2D}}\right)\right] - 1 = 0. \qquad (4)$$

In these formulae, $\gamma^* = \zeta/\zeta_{02D}$, $t_{2D} = kT/\zeta_{02D}$, $T$ – absolute temperature.

In the calculation of the electric conductivity and the Seebeck coefficient of superlattice thermoelectric material, we will assume that the dominant mechanism of free charge carrier scattering is scattering on acoustic phonon deformation potential, the free path length $l$ of charge carriers being independent of quantum numbers and inversely proportional to temperature. In so doing, we will consider the case when temperature gradient and electric field are parallel to superlattice layers.

Then the electric conductivity in model (1) will be:

$$\sigma = \sigma_{0l} \int_0^\infty \int_0^\pi \frac{y \exp\{[y + K^{-1}(1-\cos x) - \gamma^*]/t_{2D}\}}{\{\exp\{[y + K^{-1}(1-\cos x) - \gamma^*]/t_{2D}\}+1\}^2 \sqrt{2y + 4\pi K^{-2} n_0 a^3 \sin^2 x}} dxdy, \quad (5)$$

and the Seebeck coefficient is equal to:

$$\alpha = (k/e) \int_0^\infty \int_0^\pi \frac{y[y + K^{-1}(1-\cos x) - \gamma^*] \exp\{[y + K^{-1}(1-\cos x) - \gamma^*]/t_{2D}\}}{\{\exp\{[y + K^{-1}(1-\cos x) - \gamma^*]/t_{2D}\}+1\}^2 \sqrt{2y + 4\pi K^{-2} n_0 a^3 \sin^2 x}} dxdy \times$$
$$\left\{ \int_0^\infty \int_0^\pi \frac{y \exp\{[y + K^{-1}(1-\cos x) - \gamma^*]/t_{2D}\}}{\{\exp\{[y + K^{-1}(1-\cos x) - \gamma^*]/t_{2D}\}+1\}^2 \sqrt{2y + 4\pi K^{-2} n_0 a^3 \sin^2 x}} dxdy \right\}^{-1}. \quad (6)$$

In model (2), formulae (5) and (6), respectively, acquire the form:

$$\sigma = \sigma_{0l} \int_0^\infty \int_0^\pi \frac{y \exp\{[y + 2\pi^{-2} K^{-1} x^2 - \gamma^*]/t_{2D}\}}{\{\exp\{[y + 2\pi^{-2} K^{-1} x^2 - \gamma^*]/t_{2D}\}+1\}^2 \sqrt{2y + 64\pi^{-3} K^{-2} n_0 a^3 x^2}} dxdy, \quad (7)$$

$$\alpha = (k/e) \int_0^\infty \int_0^\pi \frac{y[y + 2\pi^{-2} K^{-1} x^2 - \gamma^*] \exp\{[y + 2\pi^{-2} K^{-1} x^2 - \gamma^*]/t_{2D}\}}{\{\exp\{[y + 2\pi^{-2} K^{-1} x^2 - \gamma^*]/t_D\}+1\}^2 \sqrt{2y + 64\pi^{-3} K^{-2} n_0 a^3 x^2}} dxdy \times$$
$$\left\{ \int_0^\infty \int_0^\pi \frac{y \exp\{[y + 2\pi^{-2} K^{-1} x^2 - \gamma^*]/t_{2D}\}}{\{\exp\{[y + 2\pi^{-2} K^{-1} x^2 - \gamma^*]/t_{2D}\}+1\}^2 \sqrt{2y + 64\pi^{-3} K^{-2} n_0 a^3 x^2}} dxdy \right\}^{-1}, \quad (8)$$

In these formulae, $\sigma_{0l} = 8\pi^{5/2} e^2 l \sqrt{n_0 a}/(aht_{2D})$.

The results of calculations by formulae (5-8) with regard to equations (3,4) are shown in Fig.1.

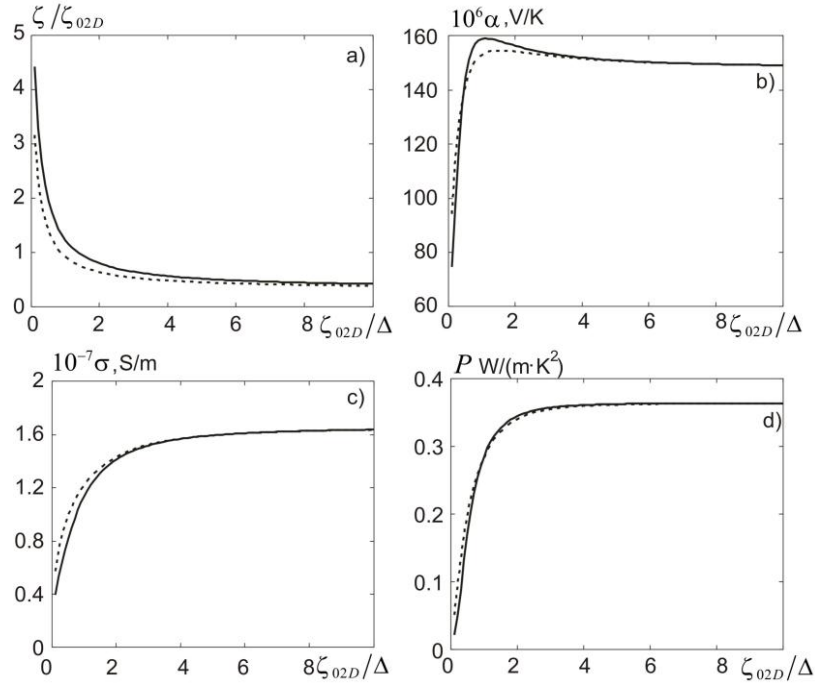

*Fig.1. Dependences on thermoelectric material Q2D degree: a) of chemical potential of free charge carrier gas; b) of electric conductivity; c) of the Seebeck coefficient; d) of power factor*

Calculations were performed with the following crystal parameters: $l = 20$ nm, $n_0 = 3 \cdot 10^{18}$ cm$^{-3}$, $a = 3$ nm and temperature $T = 300$ K.

From this figure it is evident that with increasing the Fermi surface openness degree, with constant carrier concentration the chemical potential drops, but thermoEMF, electric conductivity and power factor increase. In so doing, maximum Seebeck coefficient value occurs for transient Fermi surface for which $\zeta_{02D}/\Delta = 1$.

However, no increase in the thermoelectric figure of merit of material is observed in this case, since in conformity with the Wiedemann-Franz law, the electric conductivity growth is attended by thermal conductivity increase. Hence, quasi-dimensionality of free charge carrier gas interferes with manifestation of dimensional effects related to comparability of the electron and phonon free path lengths to film thickness or radius of powder nanoparticle. These effects should have resulted in the growth of material thermoelectric figure of merit due to the fact that in going from a single crystal to film or nanopowder material, crystal thermal conductivity is strongly restricted, and the electric conductivity – considerably less. As such, dimensional quantization of the energy spectrum does not result in the figure of merit growth, since, if we consider that the limiting values of the Seebeck coefficient and the electric conductivity correspond to "truly two-dimensional" gas of free charge carriers, and the thermoelectric figure of merit of material is an integral characteristic of a subsystem of free charge carriers in it, then the limiting dimensionless thermoelectric figure of merit of material is equal to [2]:

$$ZT = \left[\frac{(2r+5)F_{r+1.5}(\eta)}{(2r+3)F_{r+0.5}(\eta)} - \eta\right]^2 \left[\frac{(r+3.5)F_{r+2.5}(\eta)}{(r+1.5)F_{r+0.5}(\eta)} - \frac{(r+2.5)^2 F_{r+1.5}^2(\eta)}{(r+1.5)^2 F_{r+0.5}^2(\eta)}\right]^{-1}. \quad (9)$$

In this formula, $r$ -power exponent in the law of energy dependence of relaxation time, $F_m(\eta)$ – the Fermi integrals, $\eta = \gamma/t_{2D}$. Formula (9) is valid for the case of a three-dimensional free charge carrier gas with a quadratic and isotropic dispersion law. To make it valid for the case of a two-dimensional free charge carrier gas, one should replace $r$ by $r - 0.5$. Thus, in the case of constant with respect to quantum numbers free path length it should be supposed that $r = -1$. So, the limiting value of dimensionless thermoelectric figure of merit of the above material at 300K is 0.608, which is approximately equal to the thermoelectric figure of merit of bismuth telluride at this temperature [2]. Therefore, it is a good practice to use superlattice materials with a high power factor for creation of thermoelectric generators. Whereas for creation of coolers it is reasonable to use materials with a higher figure of merit whose conduction bands are closer to parabolic.

Dimensional quantization of charge carriers energy spectrum can result in the figure of merit growth, if it leads, for instance, to "semimetal-semiconductor" phase transition, as it is the case in bismuth. The mechanism of transition lies in the fact that in going from a single crystal to film or nanowire, partial overlapping of valence band and conduction band is removed, and in bismuth spectrum there appears an energy gap that radically changes the ratio between the electron and hole contributions to the Seebeck coefficient [3].

Moreover, from the results of this work it follows that superlattice materials are helpful for creation of thermoelectric devices with low electric contact resistances, as long as the electric resistance of "metal-thermoelectric material" contact is closely related to thermoelectric material resistivity.